\documentclass{ws-procs961x669}            

\newcommand{\bq}{\begin{equation}}
\newcommand{\eq}{\end{equation}}
\newcommand{\bqn}{\begin{eqnarray}}
\newcommand{\eqn}{\end{eqnarray}}
\newcommand{\nb}{\nonumber}
\newcommand{\lb}{\label}

\begin{document}

\title{Nature of Singularities in Vector-Tensor Theories of Gravity}

\author{V. H. Satheeshkumar} 

\address{Departamento de F\'{\i}sica, Universidade Federal do Estado do Rio de Janeiro (UNIRIO),\\ Rio de Janeiro, RJ 22290-240, Brazil\\
vhsatheeshkumar@gmail.com}

\begin{abstract}
The Vector-Tensor (VT) theories of gravity are a class of alternative theories to General Relativity (GR) that are characterized by the presence of a dynamical vector field besides the metric. They are studied in attempts to understand spontaneous Lorentz violation, to generate massive gravitons, and as models of dark matter and dark energy. In this article, I outline how the nature of singularities and horizons in VT theories differ greatly from GR even under the same ordinary conditions. This is illustrated with Einstein-aether theory where vacuum black hole solutions have naked singularities and vacuum cosmological solutions have new singularities that are otherwise absent in GR. It would be interesting to explore these deviations using gravitational waves.
\end{abstract}


\bodymatter

\section{Introduction}

In the 1960s and early 1970s  Roger Penrose, Stephen Hawking and Robert Geroch proved, independently and often in collaboration, a series of theorems on global features of spacetimes in GR. These theorems implied that singularities are inevitable in physically important situations of gravitational collapse and cosmology. The unpreventable existence of singularities for wide classes of rather general models in GR (and other classical theories of gravitation) marks the breakdown of the theory.  That is why understanding singularities is crucial for us to find a replacement of GR at very high energies. Even after decades of work, very little is known about the structure and properties of spacetime singularities. For the latest developments on these issues, I refer the reader to the recent reviews by Witten\cite{Witten:2019qhl}, and Penrose\cite{Penrose}, and also an accessible introduction to the subject by Joshi \cite{Joshi:2018exh}.

In this paper, we are interested in studying how the presence of a dynamical vector field present in VT theories affects the nature of singularities in comparison to GR. There are many VT theories, but of them the Einstein-aether (EA) theory has the most general diffeomorphism-invariant action involving a
spacetime metric and a vector field with the field equations being the second-order
differential equations in terms of not only the metric but also the aether field. We study both timelike and spacelike singularities in EA theory. To this end, we consider both black hole and cosmological singularities. In both cases, by computing the Kretschmann scalar, we guarantee that we are dealing with curvature singularities, not conical or caustic singularities.

The paper is organized as follows.  The Section 2 presents a quick overview of VT theories. Section 3, briefly outlines the EA theory. In Section 4 and 5, we present future and past singularities respectively. We end with a short summary in Section 6.

\section{Vector-Tensor Theories of Gravity}

One simple and straightforward extension of GR involves an introduction of a dynamical, timelike four-vector field in addition to the dynamical metric. This class of theories is referred to as Vector-Tensor theories of gravity. In some models, the four-vector is unconstrained, while in others it is constrained to have unit norm. The examples of unconstrained theories
are Will-Nordtvedt theory\cite{Will:1972zz, Nordtvedt:1972zz} (1972), Hellings-Nordtvedt theory\cite{Hellings:1973zz} (1973)  and General vector-tensor theory\cite{Will:1981cz} (1981).
The constrained theories include Kostelecky-Samuel theory\cite{Kostelecky:1989jw} (1989),  Einstein-Aether theory\cite{Jacobson:2000xp} (2001) and Khronometric theory\cite{Blas:2010hb} (2010).


The most general action for such a theory which is Lagrangian-based and whose equations of motion involving the vector-field are linear and at most of second order is given by,
\bq 
S = \frac{1}{16\pi G} \int \Big[(1+ \omega \, g_{ab}u^a u^b )R-K^{ab}{}_{mn} \nabla_a u^m \nabla_b u^n + \lambda(g_{ab}u^a u^b + 1) \Big] \, \sqrt{-g} \, d^{4}x,
\label{action}
\eq
where
\bq 
{K^{ab}}_{mn} = c_1 g^{ab}g_{mn}+c_2\delta^{a}_{m} \delta^{b}_{n} +c_3\delta^{a}_{n}\delta^{b}_{m}-c_4u^a u^b g_{mn},
\lb{Kab}
\eq
the $c_i$ being dimensionless coupling constants, and $\lambda$ is a Lagrange multiplier enforcing the unit timelike constraint on the aether. The parameter $\omega$ is taken to be zero in constrained theories, while $\lambda$ is set to zero in unconstrained theories\cite{Will:2018bme}. Below, we give the conditions under which different VT theories can be obtained from the general action given above.
\begin{itemize}
	\item Will-Nordtvedt theory: $\lambda=0$, $c_1 = -1 $ and $c_2 = c_3 = c_4 = 0$
	\item Hellings-Nordtvedt theory: $\lambda=0$, $c_1 = 2$, $c_2 = 2\omega$, $c_1 + c_2 + c_3 = 0$ and $c_4 = 0$
	\item General vector-tensor theory: $\lambda=0$, $c_1 = 2\epsilon - \tau $, $c_2 = -\eta$, $c_1 + c_2 + c_3 = -\tau$ and $c_4 = 0$
	\item Kostelecky-Samuel theory: $\omega=0$ and $u^a$ is not necessarily timelike.
	\item Einstein-Aether theory: $\omega=0$ and $u^a$ is always timelike unit vector.
	\item Khronometric theory: $\omega=0$,	$c_1 = -\epsilon$, $c_2 = \lambda_K$, $c_3 = \beta_K +\epsilon $ and $c_4 = \alpha_K +\epsilon$ where the limit $\epsilon \rightarrow \infty$ is to be taken.				
\end{itemize}

Although a lot of work is done on VT theories, there are few reviews on the subject\cite{Petrov}. From the latest phenomenological studies\cite{Oost:2018oww, Lin:2018ken, Zhang:2019iim}, it is evident that EA theory is the most representative of the VT theories and observationally relevant. For this reason, we shall restrict our discussion to EA theory. 

\section{Einstein-Aether Theory }

The general action\cite{Jacobson:2000xp} of the EA theory  is given by  
\bq 
S = \frac{1}{16\pi G} \int \Big[ R-K^{ab}{}_{mn} \nabla_a u^m \nabla_b u^n + \lambda(g_{ab}u^a u^b + 1) \Big] \, \sqrt{-g} \, d^{4}x + S_{matter},
\label{action}
\eq
where
\bq 
{K^{ab}}_{mn} = c_1 g^{ab}g_{mn}+c_2\delta^{a}_{m} \delta^{b}_{n} +c_3\delta^{a}_{n}\delta^{b}_{m}-c_4u^a u^b g_{mn},
\lb{Kab}
\eq
the $c_i$ being dimensionless coupling constants, and $\lambda$ is a Lagrange multiplier enforcing the unit timelike constraint on the aether, and 
$\delta^a_m \delta^b_n =g^{a\alpha}g_{\alpha m} g^{b\beta}g_{\beta n}.$ 
In the weak-field, slow-motion limit EA theory reduces to Newtonian gravity with a value of  Newton's constant $G_{\rm N}$ related to the parameter $G$ in the action (\ref{action}) by, 
\bq
G = G_N\left(1-\frac{c_{14}}{2}\right).
\lb{Ge}
\eq
Here, the constant $c_{14}$ is defined as
\bq
c_{14}=c_1+c_4.
\lb{beta}
\eq
Note that if $ c_{14} =0$ the EA coupling constant $G$ becomes the Newtonian coupling constant $G_N$, without necessarily imposing $c_1=c_4=0$. For $ c_{14} > 2$  the coupling constant $G$ becomes negative, implying that the gravity is repulsive. The coupling constant vanishes when $c_{14} = 2$ which renders the action undefined. Thus, physically interesting region is $0 \le c_{14} < 2$. These
free parameters have been severely constrained using many observational/experimental tests, including gravitational waves\cite{Oost:2018tcv}.

The field equations are obtained by extremizing the action with respect to independent  variables of the system. The variation with respect to the Lagrange multiplier $\lambda$ imposes the condition that $u^a$ is a unit timelike vector, thus 
\bq
g_{ab}u^a u^b = -1,
\label{LagMul}
\eq
while the variation of the action with respect $u^a$, leads to 
\bq
\nabla_a K^{am}_{bn} \nabla_m u^n + c_4 u^m \nabla_m u_a \nabla_b u^a + \lambda u_b = 0,
\eq
the variation of the action with respect to the metric $g_{mn}$ gives the dynamical equations,
\bq
G^{Einstein}_{ab} = T^{aether}_{ab} +8 \pi G  T^{matter}_{ab},
\label{EA}
\eq
where 
\bqn
G^{Einstein}_{ab} &=& R_{ab} - \frac{1}{2} g_{ab} R, \nb \\
T^{aether}_{ab}&=& \nabla_c [ J^c\;_{(a} u_{b)} + u^c J_{(ab)} - J_{(a} \;^c u_{b)}] - \frac{1}{2} g_{ab} J^c_d \nabla_c u^d+ \lambda u_a u_b  \nb \\
& & ~~~~~ + c_1 [\nabla_a u_c \nabla_b u^c - \nabla^c u_a \nabla_c u_b] + c_4 u^c \nabla_c u_a u^d \nabla_d u_b, \nb \\
T^{matter}_{ab} &=&  \frac{- 2}{\sqrt{-g}} \frac{\delta \left( \sqrt{-g} L_{matter} \right)}{\delta g_{ab}},
\label{fieldeqs}
\eqn
with
\bq
J^a\;_m=K^{ab}\;_{mn} \nabla_b u^n.
\eq
In a more general situation, the Lagrangian of GR is recovered, if and only if, the  coupling constants are identically zero, i.e., $c_1=c_2=c_3=c_4=0$.

\section{Naked Singularities in Einstein-Aether Theory }

In this section, I show that naked singularities appear in EA theory which are otherwise covered by event (Killing) horizons in GR. For this, we start with the most general spherically symmetric static metric,
\bq
ds^2= -e^{2A(r)} dt^2+e^{2B(r)} dr^2 +r^2 d\theta^2 +r^2 \sin^2 \theta d\phi^2.
\lb{ds1}
\eq

In accordance with equation (\ref{LagMul}), the aether vector is taken to be unitary and timelike,
\bq
u^a=(e^{-A(r)},0,0,0).
\eq
This choice is not the most general and is restricted to the scenario where aether is static. The aether must tip in a black hole solution as it cannot be timelike to be aligned with the null Killing vector on the horizon. As that is not the case with our choice, our solutions are valid only outside the Killing horizon. However, this is not going to be problematic for the solutions presented here.

The timelike Killing vector of the metric (\ref{ds1}) is giving by
\bq
{\chi}^a = (-1, 0, 0, 0).
\eq
The Killing and the universal horizon \cite{Wang} are obtained 
finding the largest root of
\bq
{\chi}^a {\chi}_a = 0,
\eq
and
\bq
{\chi}^a {u}_a = 0,
\eq
respectively, where ${\chi}^a$ is the timelike Killing vector. 
Thus
\bq
{\chi}^a {\chi}_a =-e^{2A(r)},
\lb{rkh}
\eq
\bq
{\chi}^a {u}_a = e^{A(r)}.
\lb{ruh}
\eq

For the metric in Eq(\ref{ds1}), the $tt$, $rr$ and $\theta\theta$ components of the vacuum field equations (\ref{fieldeqs}) are given by,
\bqn
-\frac{e^{2(A-B)}}{2r^2}\left[ c_{14}(-2 r^2 A' B'+r^2 A'^2+2 r^2 A''+4 r A') 
-4 r B'-2 e^{2B}+2\right]=0,
\lb{Gtt}
\eqn

\bq
-\frac{1}{2r^2}\left( c_{14} r^2  A'^2 + 4 r A'- 2 e^{2B} + 2\right) = 0,
\lb{Grr}
\eq

\bq
\frac{r}{2e^{2B}}\left(2 r A'' - 2 r A' B'+ 2 A'- 2 B' -  (c_{14}-2) r A'^2 \right)=0.
\lb{Gthetatheta}
\eq

The Kretschmann scalar for the metric (\ref{ds1}), is  given by
\bqn
K &=& \frac{4}{r^4 e^{4B}} \left(2 B'^2 r^2+e^{4B}-2 e^{2B}+1+2 A'^2 r^2+r^4 A''^2+\right. \nb \\
&&\left. ~~~~~~~~ 2 r^4 A'' A'^2-2 r^4 A'' B' A'+r^4 A'^4-2 r^4 A'^3 B'+r^4 B'^2 A'^2\right).
\lb{K}
\eqn

It is not possible to solve these equations analytically for a general case. However, there exist closed form solutions for $c_{14} = 16/9$ and $c_{14} = 48/25$, both of which are within the permitted range. More details about the method employed to obtain these solutions are found in our papers\cite{Campista:2018gfi, Chan:2020amr}.

For $c_{14} = 16/9$, the metric functions and Kretschmann scalar are given by,
\bqn
A&=&-\frac{3}{4} \ln(r) + \frac{3}{4} \ln{\left[r+4+\sqrt{(r+8) r}\right]},\nb\\
\\
B &=& -\frac{1}{2} \ln(2)+\frac{1}{2} \ln\left(\sqrt{\frac{r}{r+8}} + \frac{r}{r+8} \right),
\eqn
\bq
K=\frac{768 \sqrt{r+8}\left[2r(r+8)^2+(384+104 r +7 r^2)\sqrt{(r+8) r}\right]}{r^{11/2}\left(r+8 +\sqrt{(r+8) r}\right)^4}.
\eq
From the above equation, we can see that there exists a curvature singularity at $r=0$. The Killing and universal horizons for this case are obtained by finding the roots of the following,
\bq
{\chi}^a {\chi}_a =
-r^{-\frac{3}{2}}\left( r+4 +\sqrt{(r+ 8 ) r}\right)^{\frac{3}{2}}=0,
\eq
\bq
{\chi}^a {u}_a =
r^{-\frac{3}{4}}\left( r+4+\sqrt{(r+ 8 ) r}\right)^{\frac{3}{4}}=0.
\eq
We can see easily that these equations do not have any real root. Thus, there exists neither Killing horizon nor universal horizon, implying the singularity is naked. The corresponding case in GR has Killing and event horizons at $r=2M$.

For $c_{14} = 48/25$, we see the same behavior as in the previous case, but the mathematical expressions are quite lengthy. So, instead of presenting them here, I refer the reader to our paper\cite{Chan:2020amr}. 

\section{Cosmological Singularities in Einstein-Aether Theory }

In this section, I show that initial singularities appear in EA theory which are otherwise absent in GR. For this, we start with the most general isotropic and homogeneous universe is described by a Friedmann-Lema{\^{\i}}tre-Robertson-Walker (FLRW) metric,
\bq
ds^2= -dt^2+B(t)^2\left[\frac{dr^2}{1-kr^2} +r^2 d\theta^2 +r^2 \sin^2 \theta d\phi^2\right],
\lb{ds2}
\eq
where, $B(t)$ is the scale factor {and} $k$ is a Gaussian curvature of the spacial slice at a given time. The appropriate aether vector for this metric is given by,
\bq
u^a=(1,0,0,0).
\eq
The standard definitions of the Hubble parameter $H(t)$ and the deceleration parameter $q(t)$ are given by,
\bqn
H(t) &=& \frac{\dot{B(t)}}{B(t)},\\ 
\nb \\
q(t) &=& - \frac{\ddot{B(t)} B(t)}{\dot{B(t)}^2},
\eqn
where the symbol dot denotes the differentiation with respect to the time coordinate. The Friedmann-Lema\^{i}tre equations are given by,
\bqn
\left( 1 + \frac{\beta}{2}  \right) \left(  \frac{\dot{B(t)}}{B(t)} \right)^2   &=& \frac{\Lambda}{3}  - \frac{k}{B(t)^2}, \\ 
\nb \\
\left( 1 + \frac{\beta}{2}  \right)   \frac{\ddot{B(t)}}{B(t)}   &=& \frac{\Lambda}{3}, 
\eqn
where $\Lambda$ is the cosmological constant. The Kretschmann scalar for FLRW metric is  given by,
\bq
K   = \frac{12}{B^4} \left[k^2+2 k \dot B(t)^2+\dot B(t)^4+\ddot B(t)^2 B(t)^2\right].
\lb{K}
\eq
We can solve the above equations for nine different combinations of $\Lambda$ and $k$. However, here I present only the three cases involving  $k = - 1$, because these solutions are non-singular in GR but are singular in EA theory. The theoretical consistency requires $\beta \equiv c_1+3c_2+c_3 > -2$. A detailed discussion of other cases can be found in our article\cite{Chan:2019mdn}.

For {$\Lambda > 0, k = -1$}, the Friedmann-Lema\^{i}tre equations have two solutions that exist for $\beta+2 > 0$, 
\bqn
B_1(t) &=& \frac{1}{2 \sqrt{\Lambda}} \left[ -\frac{3\, e^{\sqrt{\frac{2\Lambda}{3(\beta+2)}} (t_0-t)}}{ \epsilon \sqrt{\Lambda + 3} + \sqrt{\Lambda}}  + \frac{ \epsilon \sqrt{\Lambda + 3} + \sqrt{\Lambda}}{e^{\sqrt{\frac{2\Lambda}{3(\beta+2)}} (t_0-t)}} \right],\\ 
\nb \\
B_2(t) &=& \frac{1}{2 \sqrt{\Lambda}} \left[ -\frac{3\, e^{\sqrt{\frac{2\Lambda}{3(\beta+2)}} (t-t_0)}}{ \epsilon \sqrt{\Lambda + 3} + \sqrt{\Lambda}}  + \frac{\epsilon \sqrt{\Lambda + 3} + \sqrt{\Lambda}}{e^{\sqrt{\frac{2\Lambda}{3(\beta+2)}} (t-t_0)}} \right].
\eqn
Both are singular at,
\bqn
t_{sing}(B_1) &&= t_0 - \sqrt{\frac{3(\beta+2)}{8\Lambda}} \ln \left[ \frac{2}{3}\Lambda + \frac{2}{3}\epsilon \sqrt{\Lambda (\Lambda+3)}+1 \right], \\ 
\nb \\
t_{sing}(B_2) &&= t_0 + \sqrt{\frac{3(\beta+2)}{8\Lambda}} \ln \left[\frac{2}{3}\Lambda + \frac{2}{3}\epsilon \sqrt{\Lambda (\Lambda+3)}+1 \right].
\eqn
The metric is not singular for $\beta = 0$ as the curvature invariant is $\frac{8 \Lambda^2}{3}$. But, the metric is singular for $\beta + 2 > 0$ with $\beta \neq 0$. This means it exists only in EA theory but not in GR.

For $\Lambda = 0, k = -1$, we have two solutions satisfying $\beta+2 > 0$,
\bqn
B_1(t) &=& 1 + \sqrt{\frac{2}{2 +  \beta}} (t_0 - t) \\ 
\nb \\
B_2(t) &=& 1 + \sqrt{\frac{2}{2 +  \beta}} (t - t_0)
\eqn
Both are singular at
\bqn
t_{sing}(B_1) &=& t_0 + \sqrt{\frac{2 +  \beta}{2}}, \\ 
\nb \\
t_{sing}(B_2) &=& t_0 - \sqrt{\frac{2 +  \beta}{2}}. 
\eqn
For $\beta = 0$, which corresponds to GR, the solution exists and is never singular with the curvature invariant being zero. But, the metric is singular for $\beta + 2 > 0$ such that $\beta \neq 0$. This means it is a new singularity that exists only in EA theory but not in GR.

For $\Lambda < 0, k = -1$, there exists two solutions satisfying $\beta+2 > 0$,
\bqn
B_1(t) &=&  \sqrt{\frac{3}{|\Lambda|}} \sin{\left( \sqrt{\frac{2 |\Lambda|}{3(\beta+2)}}  (t-t_0) + \sin^{-1}{\sqrt{\frac{|\Lambda|}{3}}} \right)} \,\,\,\,\,\,\,\\ 
\nb \\
B_2(t) &=&  \sqrt{\frac{3}{|\Lambda|}} \sin{\left( \sqrt{\frac{2 |\Lambda|}{3(\beta+2)}}  (t_0-t) + \sin^{-1}{\sqrt{\frac{|\Lambda|}{3}}} \right)}\,\,\,\,\,\,\,
\eqn
Both  are singular at
\bqn
t_{sing}(B_1) &=& t_0 - \sqrt{\frac{3(\beta+2)}{2 |\Lambda|}}   \sin^{-1}{\sqrt{\frac{|\Lambda|}{3}}}\\ 
\nb \\
t_{sing}(B_2) &=& t_0 + \sqrt{\frac{3(\beta+2)}{2 |\Lambda|}}   \sin^{-1}{\sqrt{\frac{|\Lambda|}{3}}}
\eqn
For $\beta = 0$ the solution exists but is never singular with the curvature invariant being $\frac{8 |\Lambda|^2}{3}$. But, the metric is singular for $\beta+2 > 0$ {with} $\beta \neq 0$. This means it is a new singularity that exists only in EA theory.

The above results are independently confirmed by studying the focusing of congruence of timelike geodesics using the Raychaudhuri equation\cite{Chan:2019mdn}.
\section{Conclusions}

Many of the deep conceptual issues related to gravity are closely connected with the very existence spacetime singularities. The singularity theorems predict only the existence of spacetime singularities under a set of physically reasonable conditions, but not say much else about them. In principle, spacetime singularities and event horizons are two totally different and independent concepts in gravitational physics. However, the Cosmic Censorship Conjecture (CCC)  suggests that whenever a spacetime singularity occurs it always stays hidden within an event horizon. But, the mathematically rigorous formulation of the CCC, let alone its proof, is still an open problem. We have attempted to understand these issues in VT theories of gravity with the hope that we might look at GR from a fresh perspective.

In this paper, we investigated how the presence of timelike vector field in EA theory affects the nature of singularities in comparison to GR theory. Firstly, I have illustrated with two analytical solutions  ($c_{14}=16/9 \textrm{ and } 48/25$) that the Schwarzschild metric in EA theory is singular at $r=0$ but it is not covered by neither Killing horizon nor universal horizon, whereas the corresponding case in GR has horizons. Thus we have naked singularities. Astrophysically, a naked singularity is distinguishable from a black hole of the same mass from the the luminosity of the accretion disk as it extends very close to the naked singularity\cite{Joshi:2013dva}. Secondly, I have shown that for three different cases with $k = -1$, the FLRW vacuum solutions in EA theory are singular, but non-singular in GR. All these solutions are within the experimentally allowed parameter space. Initial singularity is in principle visible. However, from the data we know that our universe has $k = 0$, thus making this result observationally not viable. The important take away is that the new singular solutions show that GR and EA theories have different global structures even for the simple situations.

\section*{Acknowledgments}
It is my pleasure to thank Roberto Chan and Maria de F\'{a}tima Alves da Silva for several useful discussions and collaboration on the works that produced some of the key results presented here.  I am grateful to the organizers of MG16, especially Anzong Wang, for the invitation to give this talk. Finally, I would like to dedicate this article to T. Padmanabhan (1957--2021) who has significantly influenced me before and since I got to be in his orbit when I participated in the \textit{IUCAA Vacation Students' Programme} during the summer of 2003. 


\end{document}